\documentclass[prd,12pt]{revtex4}
\usepackage{graphicx}
\begin{document}
\title{Tommy Gold Revisited: Why Does Not The Universe Rotate?}
\author{George Chapline$^{1\star}$
and Pawel O. Mazur$^{2\star\star}$
\\ $^{1}$ Physics and Advanced Technologies Directorate, Lawrence
Livermore National Laboratory, Livermore, CA 94550
\\ $^{\star}$ E-mail: chapline1@llnl.gov
\\ $^{2}$ Department of Physics and Astronomy, University of South Carolina,
Columbia, SC 29208
\\ $^{\star\star}$ E-mail: mazur@mail.psc.sc.edu}

\begin{abstract}
Understanding gravitational collapse requires understanding how
$\sim 10^{58}$ nucleons can be destroyed in $\sim 10^{-5}$
seconds. The recent proposal that the endpoint of gravitational
collapse can be a "dark energy star" implies that the mass-energy
of the nucleons undergoing gravitational collapse can be converted
to vacuum energy when one gets near to conditions where classical
general relativity predicts that a trapped surface would form. The
negative pressure associated with a large vacuum energy prevents
an event horizon from forming, thus resolving the long-standing
puzzle as to why gravitational collapse always leads to an
explosion. An indirect consequence is that the reverse process -
creation of matter from vacuum energy - should also be possible.
Indeed this process may be responsible for the "big bang". In this
new cosmology the observable universe began as a fluctuation in an
overall steady state universe. The fluctuations in the CMB in this
picture are the result of quantum turbulence associated with
vorticity. This explanation for the CMB fluctuations is superior
to inflationary scenarios because there is a natural explanation
for both the level of CMB fluctuations and the deviation from a
scale invariant spectrum at large scales.

\end{abstract}
\maketitle

\section{Introduction}

One of the oldest conundrums of cosmology is why does not the
universe rotate? Rotation and magnetic fields are ubiquitous
features in the cosmos, so it seems a little odd that rotation
does not play a role on a cosmological scale. For example, one
might wonder why rotating cosmological models such as the
G\"{o}del universe are not useful for describing the large-scale
structure of the universe. Of course, the G\"{o}del solution of
the Einstein equations represents a steady state universe, so it
would come into conflict with all the astrophysical evidence that
the universe is evolving with time. In this talk we will argue
that there is a very natural way to reconcile the evidence for a
"big bang" with the existence of rotation on a cosmological scale,
and that the failure of astrophysicists to understand the role of
rotation on cosmological scales is due to their misplaced faith in
the physical correctness of general relativity (GR) under all
circumstances.

Actually, as has been emphasized by Robert Laughlin, Emil Mottola,
and the authors in several papers over the past few years, it
cannot possibly be true that general relativity is always correct
for macroscopic length scales. In particular, in contrast with the
commonly held belief that Einstein's theory of general relativity
only fails for length scales approaching the Planck length $\sim
10^{-33}$ cm, we have argued that certain macroscopic features of
space-times that are allowed by general relativity do not occur in
the real world because they are in conflict with ordinary quantum
mechanics. The most notable of these features are event horizons
and closed time-like curves. The reason these features conflict
with quantum mechanics can be simply stated: these features are
inconsistent with the existence of a universal time based on
atomic clocks. Quantum mechanics requires for its definition a
universal time based on the synchronization of atomic clocks.
However, synchronization of clocks of any kind is not possible in
rotating space-times, and in the case of space-times containing
event horizons synchronization of clocks fails at the event
horizon.

One of us suggested some time ago \cite{GC92} that the way nature
establishes a universal time for space-time is via the existence
of off-diagonal long range order (ODLRO) for the vacuum state. In
particular, if one thinks of the vacuum state as a kind of
superfluid, then the long-range correlations between the
constituent particles of the vacuum state in effect establish a
universal time for both the vacuum state and its excitations. The
macroscopic behavior of fluids is usually described using
classical equations. However, as was first clearly explained by
Feynman \cite{Feynman55}, in the case of superfluids there are
circumstances where quantum mechanics is essential for describing
the macroscopic behavior. In the context of a superfluid theory of
space-time it turns out that these circumstances correspond
precisely to the appearance of either event horizons
\cite{CHLS00,MM01} or closed time-like curves \cite{CM04}. The
need to use quantum mechanics to describe macroscopic space-time
in these particular circumstances signals a failure of classical
GR.

The failure of a classical description of space-time near to where
GR predicts that an event horizon should occur solves a
long-standing puzzle of astrophysics; namely, how does it happen
that during the gravitational collapse of a massive stellar core
the baryon number of the core disappears in $\sim 10^{-5}$ sec?
Classical GR cannot be regarded as providing a complete physical
description of gravitational collapse because it does not tell us
what happens to the baryon number of the collapsing matter. On the
other hand, within the framework of a superfluid description of
space-time there is a direct link between the evolution of baryon
number in a region where GR breaks down and the physics of
elementary particle collisions at energies approaching the Planck
energy $\sim 10^{19}$ GeV .

According to grand unified theories of elementary particle
interactions, such as the Georgi-Glashow $SU(5)$ model, quarks can
be transmuted into leptons and mesons as a result of collisions at
energies approaching the Planck energy. The net result \cite{BC04}
is that during gravitational collapse nucleons are converted into
positrons as one approaches conditions where general relativity
predicts that a trapped surface would form. Under conditions where
the matter around the collapsing object has been ejected these
positrons can escape and form a halo around the collapsed object.
Amusingly, x-ray telescopes have detected a halo of $511$ keV
positron annihilation radiation surrounding the center of our
galaxy where a massive compact object is thought to reside.
Whether these positrons are emitted from the compact object Sag A*
remains to be confirmed, but at the present time there is no
conventional explanation for the positrons near to the galactic
center \cite{Casse}. It would be wonderful if the positron halo at
the galactic center turned out to be direct evidence for the
conversion of baryon number into lepton number predicted by Georgi
and Glashow.

\section{Destruction and Creation of Matter}

In the superfluid picture of space-time event horizons do not
occur. Instead when one approaches a condition where classical GR
predicts that a trapped surface would form, a quantum critical
layer of finite thickness forms where the red-shift becomes large
but not infinite \cite{CHLS00,MM01}. The thickness of this layer
grows as its radius increases \cite{MM01}, so that in the limit
where space-time is nearly flat the region of space in which GR
breaks down due to proximity to an event horizon becomes
macroscopically large. The behavior of matter near to a quantum
critical point is to a large extent universal, so one can infer
that nucleons passing through such a region will decay into
positrons and mesons. One can also surmise that the same sort of
process that occurs in an optical fiber where photons are
converted into a coherent squeezed state of light can occur in the
quantum critical region, and allow the energy of quark pairs to be
converted into vacuum energy. In the context of the gravitational
collapse of a stellar core this process would allow the conversion
of nucleon mass-energy into vacuum energy, and would lead to a
giant explosion since the vacuum energy has zero entropy. In
accordance with the second law of thermodynamics the entropy of
the collapsing matter must necessarily be expelled as an "entropy
exhaust".

It is tempting to identify the entropy exhaust associated with the
formation of a dark energy star with those supernovae explosions
where it has been speculated for the last half-century that a
"black hole" was formed. One problematic aspect of the black hole
hypothesis, though, has been that despite decades of effort no
explanation has ever been found to why the formation of a black
hole should lead to an explosion. Numerical hydrodynamic
calculations based on general relativity predict that when the
mass of the collapsing stellar core exceeds a few solar masses the
stellar core simply falls inside a trapped surface in a finite
proper time from which nothing can escape, and there is no
explosion. On the other hand, the dark energy star hypothesis
offers a simple explanation as to why stellar collapse always
leads to a vigorous explosion as observed from afar. Indeed, the
dark energy star prediction is that it should be difficult, just
based on looking at the visible phenomena of the collapse, to tell
whether a neutron star or a dark energy star was formed. In
particular, the visible light phenomenology resulting from the
formation of a dark energy star should resemble what happens when
a neutron star is formed, because in both cases a low entropy
residual object is formed and the entropy of the collapsing
stellar core must be removed by the ejected matter. Of course, the
physical nature of a dark energy star is quite different from a
neutron star, and this difference might be apparent in the
neutrino emissions or gravitational radiation accompanying the
supernova explosions. To date neutrino data is available for only
one supernova, $1987A$, and it is perhaps not surprising that the
results do not agree with what is expected on the basis of a
conventional picture for gravitational collapse \cite{AG03}.

The microscopic processes involved in the disappearance of
nucleons during the formation of a dark energy star are for the
most part time reversible. This means that a time-reversed process
whereby vacuum energy is converted into quarks, leptons, and gamma
rays is theoretically possible. The stability of the normal vacuum
state would preclude such a process from occurring under ordinary
circumstances. In addition, theoretical calculations indicate that
dark energy stars in isolation are stable at zero temperature. On
the other hand, gravitational collapse of an assembly of dark
energy stars may offer an opportunity to convert the vacuum energy
stored in the mass of the dark energy stars into ordinary matter
energy.

The vacuum energy density $\rho_v$ inside a dark energy star with
radius $R_{_H}$ will be given by \cite{CHLS00,MM01}
\begin{equation}
\rho_{v} = \frac{c^4}{8{\pi}G}R^{-2}_{_H}  ,
\end{equation}
\noindent where $G$ is Newton's constant. The close packing of $N$
dark energy stars, each with mass $M = {{c^2 R_{_H}}\over 2G}$,
will result in an average energy density of
\begin{equation}
\rho_{_0} = \frac{c^4}{8\sqrt{2}G}R^{-2}_{_H}  ,
\end{equation}
\noindent within a region of volume ${8NR_{_H}^3}$. This
accumulation of energy density would be gravitationally unstable
against continued collapse except for the fact that once the dark
energy stars have merged the pressure will be negative and gravity
becomes repulsive. However, the energy density (2) is far too high
for eq. (1) to be satisfied if $N >> 1$. Therefore if $N >> 1$,
almost all the accumulated mass in the merged cluster must somehow
be removed.

Eq. (1) predicts that the vacuum energy inside a dark energy star
will be $\sim 10$ times higher than the density of matter inside a
neutron star when the mass of the dark energy star is on the order
of the Chandrasekhar limit; i.e. $\sim 1.4$ solar masses. Because
the maximum mass of a neutron star is only slightly higher than
the Chandrasekhar limit, there is thus reason to suspect that the
maximum possible density for neutron stars marks a transition
between a vacuum where ordinary matter is stable and a vacuum
containing only dark energy. That is, the Chandrasekhar limit may
perhaps be interpreted as the quantum gravity Gibbs criterion for
a phase transition between nuclear matter and a state with no
ordinary matter but a large vacuum energy \cite{M03}. If this
interpretation is correct, then the transition between a state
with a large vacuum energy and a state with ordinary matter will
only occur rapidly if the merged energy density (2) exceeds by
some margin the energy density of the nuclear matter in neutron
stars. At lower merged densities the creation of ordinary matter
will not be efficient, and we would expect that the reversal of
gravitational collapse would result in an expanding cloud
consisting mainly of massive dark energy stars.

An analytical model that one might use to describe the evolution
of space-time during the conversion of the mass of a collapsing
cluster of dark energy stars into radiant energy has recently been
provided by Joshi and Goswami \cite{JG05}. In their model the
pressure varies with radius within a spherically symmetric cloud
of matter, but becomes negative in the inner part of the cloud
when the ratio of the co-moving circumference $2\pi R$ to
coordinate radius within the cloud falls below a certain value.
Initially this ratio is close to $2\pi$, but this ratio approaches
zero for all $R$ as the collapse proceeds. In order to satisfy the
Einstein equation

\begin{equation}
P = -\frac{\dot{m}(R)}{4\pi R^2 \dot{R}}  ,
\end{equation}
\noindent the mass $m(R)$ contained within a volume with
circumference $2\pi R$ must decrease with time if the pressure
$P(R)$ is negative. In the model of Joshi and Goswami the
conversion of the mass $m(R)$ into radiation is accomplished by
matching the interior metric to a Vaidya metric at the outer
boundary of the collapsing matter (cf. Fig. 1). Eventually all the
mass-energy in the collapsing cloud is converted into radiation.
This model may describe the endpoint of the collapse of a cloud of
dark energy stars in the case where the merged density (2) is much
larger than the density of neutron stars. In reality, reversal of
the gravitational collapse of a massive cloud will result in the
production of dark energy stars as well as radiation, but the
Vaidya metric will still provide a good initial description for
local space-time if the energy density post collapse is dominated
by radiation.

\begin{figure}
\includegraphics[height=7cm,width=7cm]{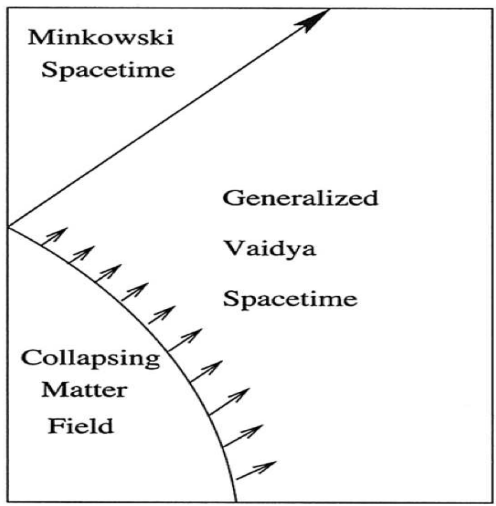}
\caption{\label{Fig1} Schematic of the behavior of space-time in
the analytical model of Joshi and Goswami that simulates the
conversion of vacuum energy into radiation.}
\end{figure}

\section{A New Version of the Steady State Universe}

In the original steady state universe of Bondi, Gold, and Hoyle it
was assumed that matter is being created at a rate that on average
is the same everywhere. It was subsequently suggested by Hoyle,
Burbridge, and Narlikar \cite{HBN00} that matter creation
processes might also lead to anomalous explosive events within
galaxies. Apparently it did not occur to them that matter creation
might be responsible for the big bang itself. Instead, they argued
that the observed evolution of the universe is an illusion. This
is not our position.

Our position is that the big bang is an illusion only in the sense
that the universe we observe may not be representative of the
universe as a whole. At the time of big bang nucleosynthesis the
entire observable universe was smaller in size than the distance
to the nearest star. On the other hand, evidence that the value of
the vacuum energy is constant with time \cite{P03} points to a
steady state universe with a scale size that is constant in time.
In particular, the coincidence between the vacuum energy predicted
by eq. (1), where $R_{_H}$ is the current radius of the observable
universe, and the value of the vacuum energy inferred from
observations of distant supernovae suggests that in the big
picture {\it we live in a quasi-steady state universe, with a
finite size on the order of $10$ gigaparsecs} \cite{MM01,M03,M97}.
Concomitantly the universe we observe may just be the long time
result of the evolution of a localized fluctuation within the
framework of a much larger universe whose overall size does not
change with time.

The explosive nature of the fluctuation leading to the expanding
and evolving universe that we observe can be naturally explained
if it is assumed that this fluctuation had its origin in the
gravitational instability of a cluster of dark energy stars. A
cluster of separate dark energy stars will not be stable against
gravitational collapse when the gravitational binding of the
cluster overcomes the expansion velocities of the stars in the
cluster. Moreover, once the dark energy stars in the collapsing
cluster begin to merge together, then the dynamics of the collapse
will change dramatically. In particular, one will evolve from a
situation where the mean pressure in the cluster is near to zero
to a situation where the mean pressure is large and negative;
resulting in a strong reversal from contraction to expansion
because the gravitational effect of negative pressure is
repulsive. If the average energy density at the time of maximum
collapse exceeds the energy density of matter in neutron stars,
then we would expect that a substantial fraction of the initial
total mass of the cluster of dark energy stars would be converted
into ordinary radiation and matter. This, of course, would match
nicely what we know about the universe that we observe, because we
know that during its initial phase the big bang was dominated by
gamma radiation and the initial density must have been high enough
to permit nucleosynthesis of deuterium and helium.

If this picture turns out to be correct, then the big bang did not
begin with infinite density, but with densities only modestly
above those that occur in neutron stars. In addition, the familiar
expansion of the observable universe would have begun at a time
only slightly earlier than the epoch when He$4$ and other light
isotopes were formed. It is perhaps disorienting that that in our
picture there is no place for all the exotic phase transitions
that have been imagined to take place in the very early universe.
In particular, there is no place for the phase transition leading
to inflation.

In our view is that there is no need for inflation; at the present
time the large scale properties of the universe are much more
easily understood in a framework where the observable universe
arises as a fluctuation within an overall steady state universe.
The problem of the lack of causal communication between different
parts of the universe in the standard big bang cosmology is
resolved by the fact that that if the observable universe arises
from gravitational collapse rather than emergence from an initial
singularity, then all parts of the observable universe would have
always been in causal contact. In particular, if one imagines that
the observable universe arose as the spherical collapse of a
cluster of dark energy stars, then during the later stages of the
collapse any point in the cluster will be able to receive light
signals from a distance $\pi R(t)$, where $2\pi R(t)$ is the
circumference of the collapsing cloud. Thus large regions of the
cluster will be able to remain in causal contact at all times
during the collapse. Initially all parts of the cluster were in
causal contact because in a steady state universe all points
within the event horizon are in causal contact. Of course, in a
strictly de Sitter universe co-moving particles always move away
from each other and eventually fall out of contact. However, this
is not true in a rotating steady state universe \cite{NST83}, and
a fortiori it is not true for co-moving particles undergoing
gravitational collapse. Thus the necessity for an inflationary
epoch in the initial expansion of the observable universe is
completely removed.

What role does rotation play if the big bang originates as a
collapsing density fluctuation in a steady state universe? As
noted earlier it is somewhat surprising that the observable
universe does not rotate. In a superfluid such as liquid helium
rotation is carried by quantized vortices \cite{Feynman55}. In the
superfluid picture of space-time rotation would also be carried by
vortex-like objects \cite{CM04}. Actually these gravitational
vortices are more like the Abrikosov vortices in a superconductor
than the Feynman-Onsager vortices in liquid helium because of the
presence of frame-dragging \cite{CM04}. In the limit where the
density of parallel spinning strings is high the spatial averaged
space-time metric approaches that of a G\"{o}del-like solution to
the Einstein equations \cite{CM04}. Since the number of vortices
will be conserved during the collapse leading up to the big bang
and subsequent expansion, the question then is why does not the
observable universe also look like a G\"{o}del universe?
Remarkably, this question has a very natural resolution in the
context of the superfluid picture for space-time \cite{CM04}. If
one starts to rotate a container of liquid helium very rapidly one
typically finds that the vortices start to meander, become
tangled, and the motion becomes turbulent. Therefore, even if the
universe as a whole rotated, in a region of space-time that was
undergoing rapid gravitational collapse the perfect alignment of
the spinning string vortices that would give rise to a
G\"{o}del-like metric will be lost, and be replaced by quantum
turbulence. Does this have any observable consequences? As it
happens quantum turbulence in a superfluid has many characteristic
features. Because there is no natural length scale in a superfluid
except for the finite size of the container, there are length
scales where the turbulence has a scale invariant spectrum. In
addition, if one assumes that the fluctuations in density are
entirely due to random variations in vorticity, then the relative
level of energy fluctuation in the vorticity field will be on the
order of $({\Delta v \over c})^2$ , where $\Delta v$ is the r.m.s.
deviation in galactic velocity from the Hubble flow \cite{M03}.

Of course, when one considers fluctuations in the CMB for the
largest length scales, the fluctuations may be expected to
remember the fact the big bang originated in an overall steady
state universe where the vorticity is not random, but a smooth
function of position as in a G\"{o}del-like universe . At these
largest scales the turbulence will be suppressed and the vorticity
will have a definite orientation in space. As it happens this is
just what is observed \cite{Jaffe05}.

\vspace{.1cm} \noindent
\textbf{Acknowledgments}

\vspace{.1cm} \noindent One of the authors (GC) is grateful to
Michael Ibison and the organizers of the Crisis in Cosmology
Conference for inviting him to participate in the conference, and
the Luso-American Foundation for financial support.

%References

\end{document}